# Offset coalescence behaviour of impacting low-surface tension droplet on high-surface-tension droplet

**Pragyan Kumar Sarma[1], Purbarun Dhar[2,b)] and Anup Paul[1,a)]**

[1]Department of Mechanical Engineering, National Institute of Technology Arunachal Pradesh, Jote – 791113, India

[2] Hydrodynamics and Thermal Multiphysics Lab (HTML), Department of Mechanical Engineering, Indian Institute of Technology Kharagpur, West Bengal – 721302, India

## Abstract

Impact of droplets of varying surface tension and subsequent spreading over a solid surface are inherent features in printing applications. In this regard, an experimental study of impact of two drops of varied surface tension is carried out where the sessile water droplet on a hydrophilic substrate is impacted upon by another droplet of sequentially lowered surface tension. The impacts are studied for different impact velocities and offsets with respect to the mid-plane of the two colliding droplets. Sodium Dodecyl Sulfate (SDS) is used to alter the surface tension without altering the viscosity, to study the various parameters affecting the spreading length viz. the surface tension, offset between the drops, and impact velocity. The spreading lengths are obtained through image processing of the captured footage of the impact dynamics by a high-speed camera. It is found out that upon lowering the surface tension, the maximum and equilibrium spreading length varies to a significant extent also the nature of the spreading dynamics changes. Both side and top-view imaging are performed to understand the overall hydrodynamics. There is also a

---

a) b) Corresponding authors: a) catchapu@gmail.com   b) purbarun@mech.iitkgp.ac.in

substantial change in "drawback" when dissimilarity is surface tension between the impacting droplets exist. Finally, a fit model is obtained to predict the maximum spread length of the various cases.

## I. Introduction

From natural occurrence to industrial usage, dynamics of droplet impacts hold significance in a wide spectrum of events and are thus a topic of widespread research and application. Starting with the pioneering works by Worthington[1], the field droplet impact dynamics have evolved largely. Typical usage of droplet impacts are in ink-jet printing, coating and spray painting, spray cooling of hot surfaces like turbine blades and rolls in metalworking, annealing processes, quenching of metal alloys, sprinklers used in fire-fighting, in IC engines, criminal forensics, fertilizer spray in agriculture, formation of salt crystals in oceans [2], etc. Understanding of droplet dynamics finds utility also in rapid prototyping[3], micro fabrication[4] and electronic packaging[5].

Studies on droplets impacting on a solid surface have shown that the most likely outcomes[2, 6-9]; are viz. deposition, spreading, receding, breakup, rebound and splashing. The aspects that influence such regimes are the velocity of impact, size of the droplet, surface tension, surface temperature, viscosity, wettability, ambient pressure, etc[10, 11]. Chandra and Avedisian[12] studied the dynamics of droplet impact on a thin lamina of liquid for different surface temperatures and proposed a model for the spreading kinetics. Pasandideh and Fard[13] and Fukai et al.[14] predicted the maximum spreading length of a droplet numerically via a modified Volume of Fluid approach. Experimental and theoretical finding on multiple droplets impacting a solid substrate was reported by Roisman et al.[15]. Their model for the maximum spreading factor accounted for the effects of inertia, surface tension, viscosity and surface wettability. Park et al.[16] discussed a scaling model for maximum spreading factor in correlation with the Reynolds number and Weber number and validated it for millimetre and micron-scale droplets. Various impact outcomes were reported for droplet impact on hydrophobic or superhydrophobic surfaces[6, 17, 18]. Sahoo et al.[19] investigated the role of the inclination of the surface on post-impact dynamics of drops and discussed the roles of dimensionless numbers, inclination level and wettability.

[Type here]

Inkjet printing involves rapid deposition of tightly spaced droplets on a substrate and the final print quality is governed by the deposition and coalescence efficiency of the neighbouring droplets [20, 21, 22]. Droplet coalescence behaviour finds prominence in fields like printing of 'smart' materials, biomaterial deposition, production and treatment of pharmaceuticals, etc[23-26]. Some important parameters that govern the quality of deposition or printing are spacing between two consecutive droplets, the time gap between the droplets, the surface energy of substrate, fluid properties, etc. Duniveld[27] experimentally observed that the non-uniformity of a printed line's thickness is created by the development of a series of liquid blisters joined by liquid ridges during the droplet coalescence process. Li et al[28] studied the impact of molten wax drops on a solid substrate, which were deposited along a straight line, and then solidified. They reported a phenomenon called 'drawback', which describes the pulling action of the second falling droplet onto the first deposited droplet, caused by the surface tension between two overlapping droplets. Such events lead to breaks in the continuity and homogeneity of deposited or printed structures.

The coalescence dynamics of droplets on a wetting surface has been studied by Ristenpart et al.[29] and they discussed the width of the evolving meniscus bridge between the drops. Lee et al.[30] studied the coalescence of droplets of different viscosities and noted its effect on the dimensionless spreading length. They found that the larger viscosities contribute to the lesser breakup of the deposited drops. The spreading dynamics and oscillations of a single, and subsequently deposited droplets on surfaces of varied wettability have been studied by Yang et al.[31]. The dependence on spacing, deposition frequency and time scale of droplet evaporation with the structure of coalesced drops and the arrangement of suspended particles due to drying of such deposited droplets has been discussed by Yang et al.[32]. A probe on nano colloidal droplet deposition on porous structures has been carried out by Chiolerio et al.[33], where they discussed the jetting effects characterised by droplet spacing and frequency of ejection on determination of the arrangement of nanoparticles in reactive ink droplets.

Graham et al.[34] reported experimental and numerical work on coalescence dynamics on substrates of various wettability and developed correlations for the prediction of maximum spreading length. Experiments by Li et al[35] shed light on the criteria for the spread length of coalescing droplets being larger or smaller than the ideal spread length. They identified three different coalescence dynamics based on the maximum and minimum spread lengths and developed correlations to determine conditions for the formation of continuous deposition or print lines. Kalpana et al.[36] studied the coalescence and spreading of conducting polymer droplets on substrates of varied wettability, for different offset impacts, and compared the different effects concerning head-on impact of the second droplet. They identified different governing forces in different regimes of spreading with the help of an empirical formulation. The internal dynamics during coalescence of two drops with distinct surface tension were reported by Sykes et al.[37]. They use both the side and bottom views to highlight the internal dynamics of the coalescence, and report that a jet formation occurs on top of the coalesced drops which is a result of the surface flow due to inertia and contact line immobility. This jet can be accentuated or reduced by the surface tension contrast of the drops.

The literature shows that a fundamental understanding of the coalescence dynamics of two droplets, at various offsets of impact, and with pre-defined surface tension contrast between them is of immediate significance towards several utilities as discussed above. Being aware that no such works have been done previously, we experimentally study the coalescence dynamics of a low surface tension fluid droplet impacting on a higher surface tension fluid droplet (rested on a glass substrate). We consider a water droplet as the initially seated droplet and use surfactant solutions of different concentrations to obtain the second droplet of lowered surface tension. The use of surfactant also ensures that the viscosity of the fluid changes only very minorly with respect to water. The impact of the second droplet is studied at different offsets with respect to the central vertical axis of the initial sessile droplet. Also, the role of the impact velocity is studied by allowing the droplets to free fall from different altitudes. A water drop falling on a water drop, for various offsets and velocities, has been considered as a control. We analyse and discuss the various fluid dynamics aspects associated with such surface tension difference driven droplet coalescence for various offsets and impact conditions.

## II. Materials and methodology

The experimental setup (schematic in Figure 1(a)) comprises a droplet dispensing glass microliter syringe (Gilmont instruments, 2.0 ml) mounted on a height-adjustable holding fixture. The droplet is pumped out by an electronically controlled dispenser (Ramé-hart instruments, USA) and carefully deposited on a glass slide (thoroughly cleaned with acetone and DI water and dried in a hot air oven). The slide with the sessile droplet is carefully positioned below the mounted glass syringe on a 3-axis movable stage. DI water is used as the initial sessile droplet, and droplets

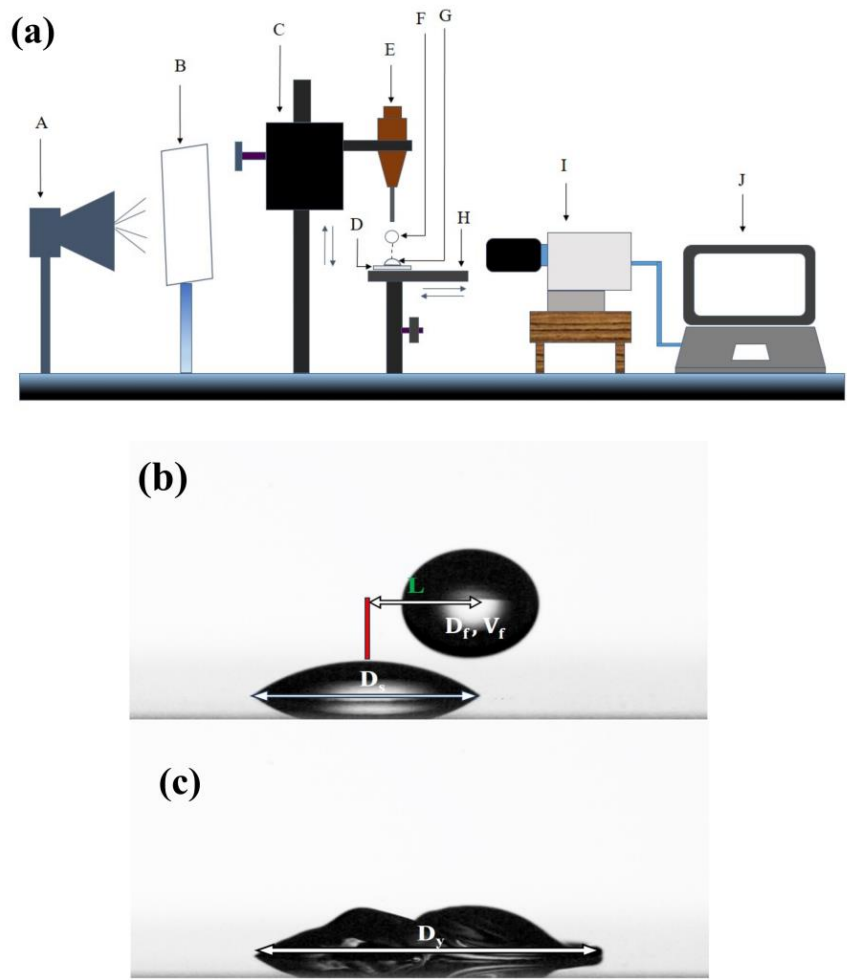

**Figure 1: (a) Schematic representation of the experimental set-up (A) light emitting diode (LED) assembly (B) light diffuser plate (C) height adjustable syringe mounting assembly (D) glass slide (E) microliter syringe (F) falling droplet (G) sessile droplet (H) horizontally movable stage (I) high speed camera (J) computer system. (b, c) Pictorial illustration of the parameters of droplet falling on another drop. (b) Indicates the dimension of the falling drop, stationary drop, the offset distance and the falling drop velocity. (c) Indicates the spreading length of the post coalescence**

of varied surface tension are allowed to impact it at velocities of 0.4 m/sec, 1 m/sec and 1.64 m/sec by adjusting the height of free fall. The offsets are set by horizontally moving the specimen stage (accuracy of ~ 1 μm). Photography was done (at 3600 frames/sec, resolution of 1280 x 800 pixels for both front and top views) using a high-speed camera (VEO 340 Phantom) fit with a macro lens (Nikon) with LED illumination. Images were. The images were post-processed with the aid of the open-source software FIJI (ImageJ+). Test fluids of different surface tension (with only insignificant change in viscosity) were aqueous solutions of sodium dodecyl sulphate (Sigma-Aldrich) in DI water. The concentrations studied were 0.25, 0.5, 0.75, and 1 CMC (Critical Micelle Concentration). The surface tension and viscosity were measured by pendant drop analysis (Ramé-hart instruments, USA) and coaxial cylinder rheometer (Brookfield, U.S.A) respectively, and are tabulated in Table I.

**Table I:** Properties of the test fluids at 25°C. All experimental variations are correct to within ±2% of the mentioned values.

| Liquid | Surface tension (N/m) | Viscosity (mPas) | Avg. diameter of falling droplet (mm) |
|---|---|---|---|
| DI water (0 CMC) | 0.071 | 0.89 | 2.52 |
| SDS 0.25 CMC | 0.045 | 0.89 | 2.46 |
| SDS 0.5 CMC | 0.032 | 0.89 | 2.39 |
| SDS 0.75 CMC | 0.025 | 0.90 | 2.36 |
| SDS 1 CMC | 0.020 | 0.90 | 2.25 |

In this study, we have experimented on the impact of the second falling droplet on the first sessile droplet, for various impact velocities and offset between the centre lines of the two droplets. Figure 2 shows the representative images of droplet spread, offset, and non-dimensional time, and their expressions are as follows (Li et al.[35] and Sarojini et al.[36])

$$\Psi = \frac{D_y}{D_s + L} \quad (1)$$

$$\lambda = 1 - \frac{L}{D_s} \quad (2)$$

$$\tau = \frac{tV_f}{D_f} \qquad (3)$$

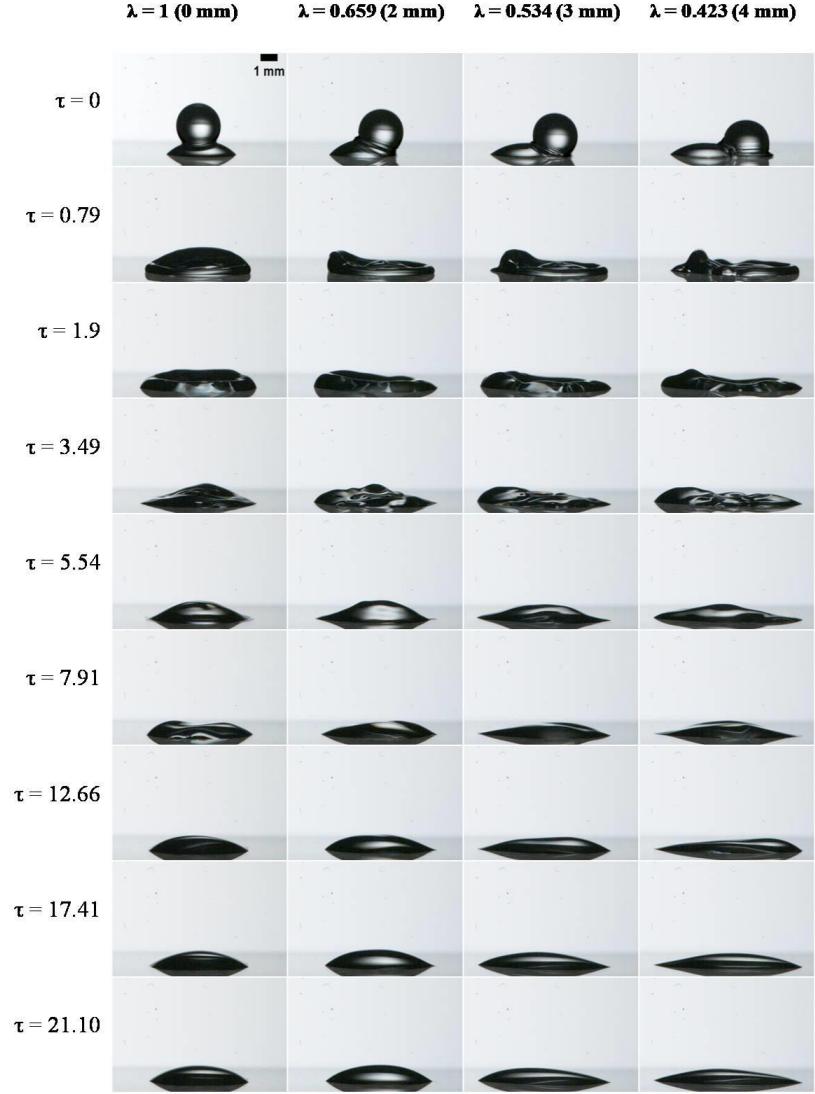

**Figure 2: Time series snapshot of a water droplet colliding at a velocity of 1 m/sec with a stationary water drop at various offsets (λ).**

Here, $\Psi$, $\lambda$, and $\tau$ are the non-dimensional spreading length, overlap ratio or non-dimensional offset length, and non-dimensional time, respectively. $D_y$ is the instantaneous spreading length of the post-coalesced structure, $D_s$ is the diameter of the initial sessile droplet, $L$ is the centerline offset between the sessile drop and the falling drop, $t$ is the time elapsed after the second drop first touches the first drop, and $V_f$ is the impact velocity. Some non-dimensional numbers considered in this study are $Re = \rho V_f D_f / \mu$, $We = \rho V_f^2 D_f / \sigma$, $Ca = \mu V_f / \sigma$ where Re, We, Ca denotes Reynolds number, Weber number and Capillary number respectively. $\rho$ is the density of the fluid and $\mu$ is the dynamic viscosity of the fluid.

## III. Results and discussions

### a. Dynamics at impact velocity of 1 m/s

Figure 2 illustrates the post-impact dynamics and variation of the spreading length for a water drop falling with a velocity of 1 m/sec (We = 58 and Ca = 0.012) on another sessile water drop. The corresponding top views are shown in Fig. 3. Figs. 2 and 3 serve as a control where the surface tensions of both droplets are the same. Fig. 4(a) shows the variation of non-dimensional spreading length $\Psi$ with respect to non-dimensional time $\tau$. It can be inferred that the head-on collision makes the coalesced droplet reach to a significant level owing to the exertion of maximum impact inertia. The same goes for droplets falling at different offsets but their maximum spread is not as significant as that of the head-on collision as a result of lesser collision energy on the sessile drop. In the case of head-on collision ($\lambda=1$), there is a strong pullback of the surface tension forces as the impact energy depletes ($\tau = 3.49$ Fig. 2). The pulling action results in overshooting the equilibrium length and also due to surface pinning of the edges results in re-spreading spreading after receding as can be seen in the Fig 4(a) after $\tau = 5$. There is an oscillation due to re-spreading and pullback action and the oscillations get dampened out as time progresses and finally the blob of liquid reaches an equilibrium configuration. The maximum spreading length decreases as the offset length between the droplets increases which can be observed from Fig 4(a) owing to the less impact energy transmitted on the sessile drop by the falling one. There is a delay in reaching the maximum spread length at higher offset values ($\lambda = 0.534, 0.423$) because more time is required for the non-impacting end of the static droplet to set into motion as the pressure pulse is opposed by the interfacial tension[36]. Upon reaching the maximum spread length the impacting edge retracts faster than the non-impacting edge which is already explained in the above line, the non-impacting end takes more time to receive the momentum transfer and by that time the impacting end spreads and due to viscous effects caused by the dry glass substrate as well as due to the surface tension of the liquid and retracts a bit and gets pinned to the surface. This pinning action caused internal redistribution of fluid and a bulge to form at the centre and move towards the non-impacting end and finally, after a passage of time the coalesced liquid attains an equilibrium length. It is also quite evident from Fig 4(a) that on larger offset droplet impact ($\lambda = 0.534, 0.423$) the retraction length is quite low in comparison to droplet impact at head-on and smaller offset ($\lambda = 0.659$) and the equilibrium length nearly reaches the maximum spreading length, all these are due to the pinning action of the edges.

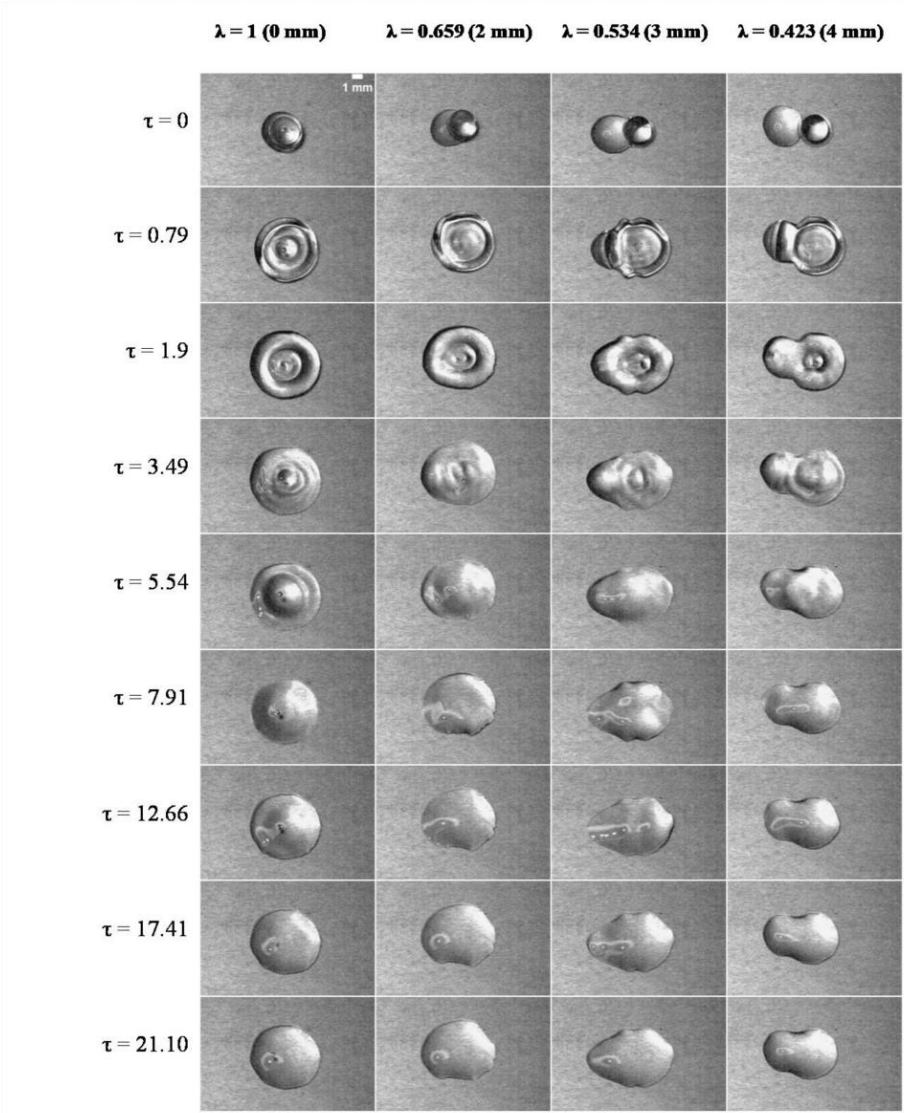

**Figure 3: Time series snapshot of top views of a water droplet colliding at a velocity of 1 m/sec with a stationary water drop at various offsets (λ).**

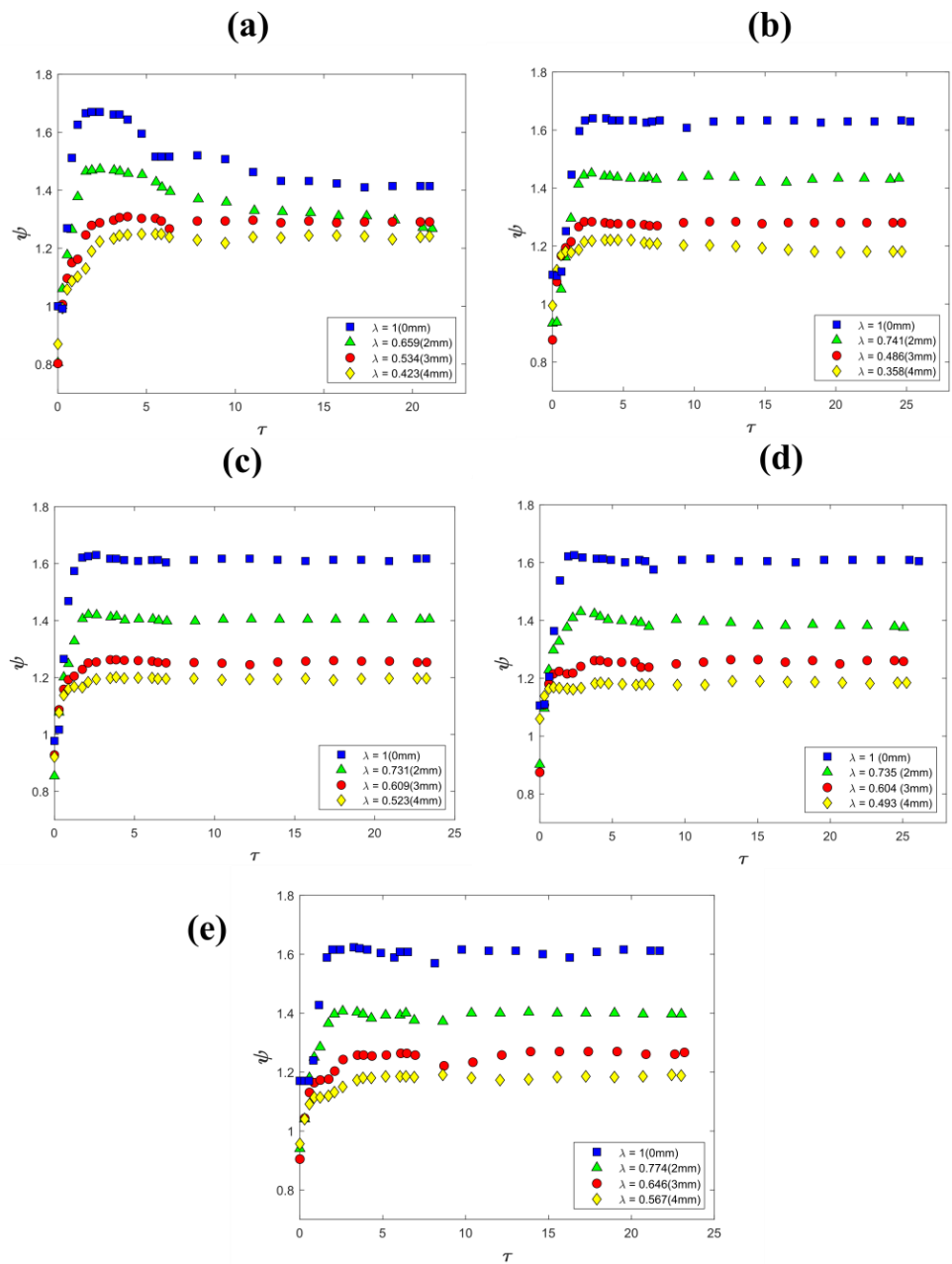

**Figure 4: Non-dimensional spread length (ψ) vs. non-dimensional time (τ) at various offsets (λ) at 1 m/sec velocity for (a) Water drop falling on water drop (b) 0.25 CMC aqueous solution of SDS on water drop (c) 0.5 CMC aqueous solution of SDS on water drop (d) 0.75 CMC aqueous SDS solution on water drop (e) 1 CMC aqueous solution of SDS on water drop.**

The top views of the droplet spread illustrated in Fig. 3 gives a comprehensive idea of how the lateral and asymmetric spreading occurs. The topography of the composite drop at the maximum spread at each different offset can be visualized from the top view images. At $\tau = 1.9$ (Fig. 3) the maximum spread occurs for λ = 1 and 0.659 (head-on and 2 mm offset) and it is clear that the head-on impact creates a rim-lamella spreading and it is symmetric spreading increases. The higher offset impacts at λ = 0.543, 0.423 the maximum spreading length reaches at τ = 3.49 since the pressure pulse from the impact of the impacting drop reaches the non-impacting end of the sessile drop overcoming the resistance offered by the sessile drop owing to the interfacial surface tension. The top view images also show that at the highest offset the composite droplet starts forming a line whereas at the lower offsets, the compound drop forms a puddle.

The falling droplet (v= 1m/sec, We = 78, Ca = 0.019, Re = 4074) is changed to an aqueous solution of sodium dodecyl sulfate of lower concentration (0.25 CMC), this results in lowering the surface tension of the falling droplet to that of the sessile droplet and keeping the viscosity of both the falling and sessile droplet unchanged. Figure S1(refer Supplementary material) represents the snapshots and Figure 4 (b) shows the graphical depiction of the non-dimensional spreading length against non-dimensional time. The falling 0.25 CMC SDS solution droplet hits the sessile drop head-on and at three different offsets. The head-on collision makes maximum impact on the sessile water drop and the compound drop deforms to a pancake shape making a depression in the middle causing the compound drop to form a toroidal shape and spreading outwards equally in both directions. The maximum spread length occurs at around τ = 2.21 and as the impact energy dissipates against viscosity and surface tension the minimal surface tension does not aid in much pull-back. Additional spread due to Marangoni effects cannot be evaluated from the current work. The spreading edges get pinned onto the glass surface earlier than what was observed on water-on-water impact (Fig 2), this results in creating a bulge at the centre at τ = 6.41. The bulge subsequently disperses uniformly and the compound drop flattens out due to the pinning action from contact angle hysteresis, this is in contrast to what was observed on the water drop impacting a water drop in Fig.2 (λ = 1). The smaller offset (λ = 0.731) impact results in a non-axisymmetric shape as the right edge feels the disturbance a little later because the static droplet has higher surface tension and the lower surface tension fluid hasn't mixed uniformly. After the compound droplet flattens out the receding action is quite similar to the head-on impact case. The larger offset impacts i.e., λ = 0.486 has a part of the falling droplet land on the static one and a part lands in the solid substrate, in λ = 0.358 lands on the solid substrate and spreads and mingles with the static drop. The impact of the falling drop pushes the static drop and spreading starts, in both these cases the surface wave travels to the non-impacting edge due to interfacial tension after that there is a relaxation and pinning of the edges occurs at around τ = 6.41. The fluid inside the compound drop redistributes and finally settles into its equilibrium length.

Fig. 4(b) gives the quantitative aspect of the non-dimensional spread with respect to non-dimensionless time, it is clear from the figure that the head-on impact results in maximum spread length but as the inertial regime depletes there is not a significant pullback like the one seen in the previous result (Fig 4(a)). It is interesting to notice that the behaviour is similar to the one where the impacting droplets' viscosity is increased[30]. For the head-on case, the time required to reach the maximum spreading length is almost equal to the one achieved in the water-on-water case. The maximum spreading length decreases as the offset between the droplets increases which is obvious as the impact inertia partially reaches the static droplet and the static droplets' surface tension acts as a resisting medium. The maximum spreading length of the larger offset impacts of the 0.25 CMC SDS solution is lower than the water-on-water impact case because the surface pinning happens earlier when the droplet surface tension is lowered. Figure S2 (refer Supplementary material) represents the top views of the 0.25 CMC droplet impacting the sessile water drop. The head-on collision makes the composite droplet spread out uniformly in all directions. In the concentric impact case, the coalesced droplet's retraction happens from τ = 6.41, the composite droplet doesn't retract uniformly as compared to that of a water droplet falling on a water drop although the retraction is slight (observed from Fig. 4 (a)) when surface tension is lowered to 0.25 CMC it is mainly due to non-uniform mixing of both the unsimilar surface tension liquid drops. The higher surface tension of the sessile droplet in this case offers a larger interfacial tension resistance to the impacting drop as it spreads upon impact and it is more pronounced when the offset value increases. From Fig S2 at the highest offset ratio of λ = 0.486, 0.358 the left end of the sessile droplet feels the disturbance later time than the one with lesser offset (head-on and 2 mm offset). For the largest offset case, it is observed from the top view that a fine line is formed instead of a puddle which happens in the case of all the lower offset impacts.

Figure S3 (refer Supplementary material) and Figure 4 (c) describes the impact of a falling droplet of aqueous solution of sodium dodecyl sulfate (0.5 CMC) falling (v = 1 m/sec, We = 117, Ca = 0.027, Re = 4331) on a sessile water droplet concentrically and at different overlaps. The post-impact dynamics of a head-on collision and at a lesser offset distance (λ = 0.731) show similar dynamics as discussed for the case of 0.25 CMC aqueous solution hitting a water droplet. The falling droplet impinges the sessile drop with maximum inertia and pushes out the static droplet radially in all directions. The maximum spreading for concentric collision occurs at τ =

2.612 which is nearly about same for the case of water-on-water impact and drop and 0.25 CMC aqueous solution of SDS falling on water. Due to the lowering of surface tension, there isn't much drawback for all different edge collisions, at $\tau = 6.14$ the composite drop's edge gets fixed and there's a bulge in the centre for the head-on collision and towards the non-impacting side for 2mm offset collision ($\lambda = 0.731$) after which the surface curvature decreases towards an equilibrium position. The larger offset ($\lambda = 0.609, 0.523$) impacts also show similar dynamics as discussed in the previous case. Impacting inertia causes the impacting fluid to spread and touch and push the static drop (water) which is at higher surface tension. The pressure pulse is resisted by the interfacial tension of the static droplet and there is a deformation of the free surface. At $\tau = 3.87$, the impacting edge gets pinned for both the larger offset impacts because there will be a pullback due to surface tension gradients towards the impacting ends which would be resisted by viscosity. The non-impacting edge also gets pinned at $\tau = 6.14$ which causes the bulge to move towards the non-impacting side and finally settle to an equilibrium position at $\tau = 23$. Fig 4(c) gives the quantitative analysis of the spreading length against time. The head-on impact shows the maximum spreading which is agreeable because the static droplet gets the maximum impact inertia. Maximum spreading length is reached at $\tau = 2.61$ which is roughly the same as the 0.25 CMC aqueous SDS solution droplet falling on the water drop case. After the steep inertial regime there is a weak capillary regime where the edges recede to a very lesser extent which is very similar to a 0.25 CMC aqueous SDS solution falling on a water drop beyond the weak capillary regime the combined blob doesn't spread much although the competition of the inertial and capillary forces doesn't die out there are still some minor oscillations which eventually becomes stable and the equilibrium length becomes almost the maximum spreading length. For the minor offset ($\lambda = 0.731$) we see that there is significantly less maximum spreading length than the concentric impact although it reaches the maximum spread length at roughly the same time taken ($\tau = 2.63$) for the head-on impact. Due to the reduction in surface tension of the falling droplet, the receding regime is short-spanned. The larger offsets ($\lambda = 0.609, 0.523$) attain the maximum spreading length a little later than the smaller offset and head-on impact which is $\tau = 3.487, 3.847$ for $\lambda = 0.609$ and $\lambda = 0.523$ respectively. The larger offset impact shows negligible receding and the spreading plot shows a similar nature to a highly viscous drop impacting on another drop[30]. It means there is no further spreading after the maximum spreading length is received. This type of spreading is favourable for line printing where the spreading and drawback phenomena are not desirable.

From the top views of the same dynamics can be seen in Figure S4 (refer Supplementary material), the concentric collision makes a uniform circular spread, retraction occurs from $\tau = 6.14$ and it is also observed from the front view images that upon lowering the surface tension of the impacting droplet the capillary rise height gets lowered during retraction as a result of contact line pinning. The concentric symmetric spreading vanishes when the falling droplet impacts at the lower offset ratio of $\lambda = 0.731$ (2mm) which means even a small lateral separation makes the composite drop spread in an oblong shape, the travelling capillary wave towards the non-impacting side at $\tau = 2.12$ for the offsets of $\lambda = 0.731, 0.609$ which lends credence that the static droplet act as a resistant to the inertial force of the impacting droplet due to its higher surface tension, a fact that can be observed for the highest offset ratio of $\lambda = 0.523$ (4 mm) where the non-impacting end of the sessile droplet remains undisturbed even at $\tau = 6.14$. For the lateral separation of 3mm and 4 mm, the spreading on the non-impacting side is minimal and it is the impacting side that spreads the most and lowering of surface tension as well as the hydrophilic nature of the substrate aids in that extent of spreading. The formation of a line could be seen from the highest offset impact ($\lambda = 0.523$) although the impacting edge gets more spread in the longitudinal direction making it a dumbbell-shaped line.

On increasing the concentration of SDS so that the aqueous solution reaches 0.75 CMC to lower the surface tension further without altering the viscosity and letting a droplet fall on a sessile water drop we get the impact and spreading dynamics as featured in Figure S5 (refer Supplementary material) and Figure 4(d). The head-on impact creates an axisymmetric crater-like dent due to the highest transfer of impinging inertia on the static water drop. The kinematic phase of spreading extends till the compound drop reaches a maximum spread length at $\tau = 2.366$ around the same time that is required for 0.5 CMC and 0.25 CMC compound droplet required to reach after head-on impact. There is a receding phase which takes about 11.94 ms, the receding phase happens because due to the sharp difference in surface tension, the lower surface tension fluid just rests over the static droplet and it takes a longer time to mix[37] and the droplet retracts back due to the surface tension of the static droplet. Due to the capillarity, there is some re-spreading after receding and after $\tau = 9.79$, the spreading length remains the same albeit with some minor oscillation and the equilibrium spreading length remains the same as the maximum spreading length. The smaller offset ($\lambda = 0.735$) impact shows a similar trend as the previous lower surface tension fluid impact showing it reaches the maximum spreading length at a similar time frame as that of the concentric collision and there is a receding phase as well which lasts about 16.6 ms. The higher offsets dynamics also show similar dynamics as discussed earlier but the difference is that there is a relaxation in spreading for a short duration as the surface waves propagate from the impacting towards the non-impacting end as seen on λ = 0.604 at τ = 2.27 (Fig. 4 (d)) then the bulk fluid reach a maximum length at τ = 3.757, the recoiling wave bulges out and goes towards the non-impacting end and finally towards the centre and settles for an equilibrium position. The largest offset (λ = 0.493) impact attains the maximum spreading length at the longest time which is intuitive because the falling droplet lands on the solid glass surface and spreads radially at first and then the radial spread pushes and mixes with the static droplet after which the combined droplet merges and starts spreading as a whole. There isn't much spreading happening after the maximum length is reached with the farthest offset impact which more-or-less behaves as a high viscous droplet impact[30]. The kinetic energy depletes on pushing the static droplet and against viscosity also the lower surface tension isn't pulling back the impacting edge so there isn't much receding altogether. There is a minimal bulging towards the centre and the fluid just redistributes over the surface and reaches to a static state. The overhead images of the impact dynamics of the 0.75 CMC droplet on the water droplet are shown in Figure S6 (refer Supplementary material). The equilibrium shapes of the head-on impact and the lower offset impact (2mm) are quite similar to the one described in the preceding droplet impact. The shape at maximum spreading for the head-on collision is symmetric whereas for a small offset, the symmetrical shape distorts and forms a puddle at the equilibrium spreading length. For the 3mm offset ($\lambda = 0.604$) the left edge feels the disturbance at $\tau = 4.14$ which is later than the previous smaller offset impact (head-on, 2mm) hence the maximum spreading happens towards the impacting side which is similar to the largest offset impact as well and this phenomenon is observed in the previous droplet impacts. The equilibrium shape of the largest offset impact ($\lambda = 0.604$) makes a line but the larger radius spreading of the impacting drop due to the inertial effect makes the composite drop take the shape of a dumbbell as well. The least surface tension that could be attained without changing the viscosity significantly is by reaching the critical micelle concentration by adding surfactants. So now the least surface tension fluid droplet which is let fall on the static water droplet is 1 CMC of aqueous SDS solution.

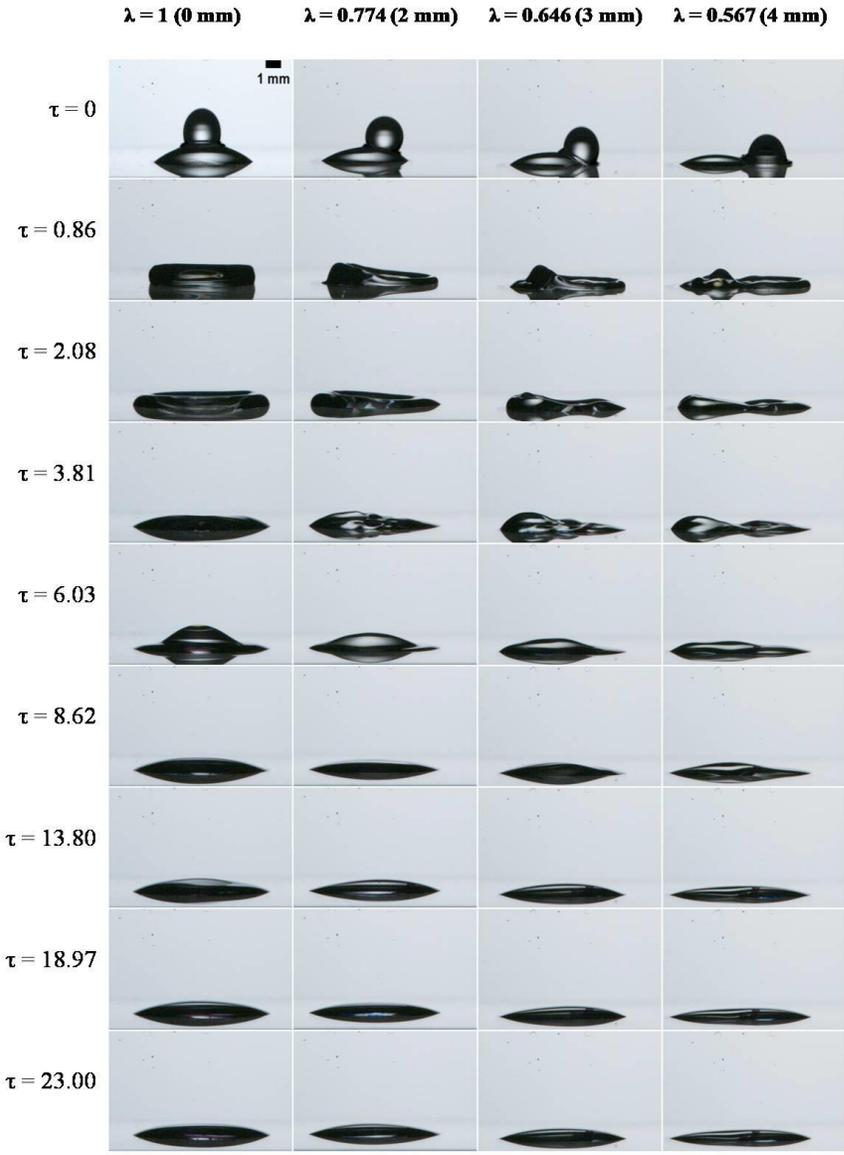

**Figure 5: Time series snapshot of a SDS solution of 1 CMC colliding at a velocity of 1 m/sec with a stationary water drop at various offsets**

The post-impact and spreading dynamics both qualitatively and quantitatively are shown in Figure 5 and Figure 4(e). The head-on collision leads to the maximum transfer of kinetic energy onto the static droplet causing the compound drop to spread in a concentric rim-lamella manner which expands causing the rim height to decrease and reach the maximum spread length at $\tau = 3.25$ followed by retraction for a period of 1 ms , the composite drop re-spreads and the edges get fixed at $\tau = 8.62$ and there is minimal spreading however due to surface tension the free surface bulges and finally settles down to an equilibrium state. The dynamics of the lesser offset are similar to the head-on impact except the impact is not axisymmetric, the lamella from the right side moves towards the right reducing in height, Fig 4 (e) shows that the maximum spreading length occurs at the same time as the concentric impact which is $\tau = 2.58$, there is a short burst of retraction seen here as well and followed by re-spreading and finally, the equilibrium spreading length is closer to the maximum spreading length is also the same with the head-on impact. The larger offset λ = 646 shows the same dynamics as the 0.75 CMC droplet falling on the water droplet the maximum spreading length is reached later than the head-on and 2 mm (0.774) offset impact at τ = 3.47, there's a retraction phase from τ = 6.08 to 8.69 and the fluid respreads reaches an equilibrium length which is also the same as the maximum spreading length. The largest offset impact is for λ = 0.567, the droplets impact the solid substrate and then spread like a single drop impact, the radial jet impinges on the sessile water drop and the surface wave propagates from the impacting end towards the non-impacting end causing the compound drop to spread outwards, the impacting edge pins earlier than the non-impacting end causing a bulge to appear on the non-impacting end and it moves towards the right side.

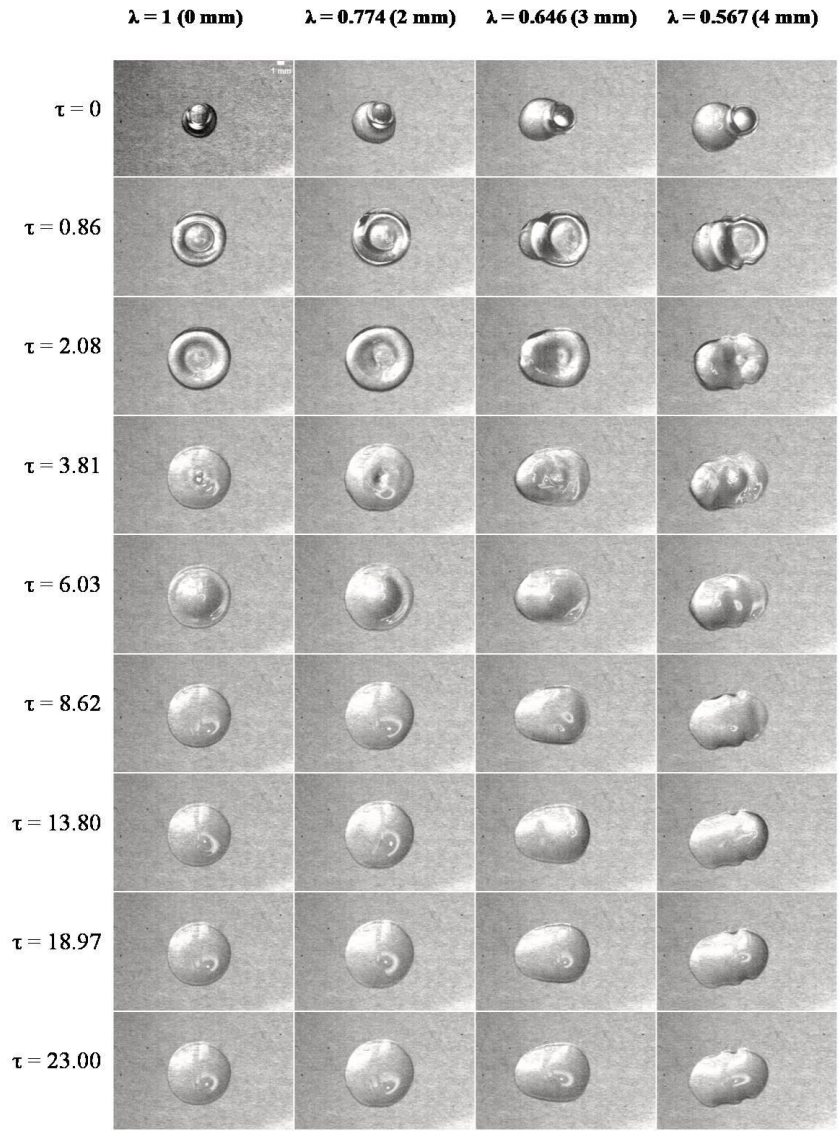

**Figure 6: Time series snapshot of top views of a SDS solution of 1 CMC colliding at a velocity of 1 m/sec with a stationary water drop at various offsets**

The maximum spreading length is reached at τ = 3.8, followed by a gradual retraction in spread length, the retraction is dissipated through viscous action and the capillary action leads to re-spreading which settles for an equilibrium length that is roughly the same as that of the maximum spreading length. Figure 6 elucidates the impact dynamics of the same as seen from the above. Concentric and smaller offset collision spreading happens in a similar manner as discussed previously. On increasing the offset (λ = 0.646, 567) the spreading dynamics change as there is a more longitudinal spreading which creates a more puddle-like shape even for the largest offset impact. For the largest offset of 4 mm (λ = 0.604), the necking region gets a lamella protruding longitudinally which creates the equilibrium length to be a more tick line rather than a fine line which happened on higher surface tension impacting droplet (0.25 CMC, 0.5 CMC). Hence it is seen that lowering the surface tension of the impacting droplet leads to a bleeding-like effect which prevents the formation of fine lines.

*b. Dynamics at impact velocity of 0.4 m/s*

The spreading length against non-dimensional time of the same fluids at the same offsets at a lesser velocity of 0.4 m/sec further highlighting the effects of surface tension are illustrated in Figure 7(a-d). The needle height of the syringe was lowered for the droplet to attain a velocity of 0.4 m/sec. The Weber number ranges from 8.5 – 26, Capillary number ~ O(-3), Reynolds number varies from 1506 – 1738. As discussed previously the falling droplet is varied in surface tension ranging from water which is treated as a controlled experiment where the sessile drop and the falling drop have the same fluid properties to lowering the surface tension of water by adding surfactants where the concentrations are varied with respect to percentage of the critical micelle concentration (CMC) as discussed previously. The impacting droplet is allowed to fall on the sessile droplet concentrically and also at some lateral separations varying from 2mm, 3mm and 4 mm. off the centre of the sessile drop.

Figure 7 (a) shows the non-dimensional spread length vs. non-dimensional time of a water drop falling on a water drop at the centre and at different offsets. The height from which the falling droplet is squeezed out is low so the falling drop oscillates and its shape doesn't always stay spherical as it hits the sessile water drop. The low velocity of ejection is indicative that the inertial energy is low, hence the falling droplet is unable to create a dent on the static drop and instead gets squeezed there is a likelihood that the air film trapped between the droplet doesn't get drained out much which could result in cushioning of the falling droplet over the sessile one. As the falling

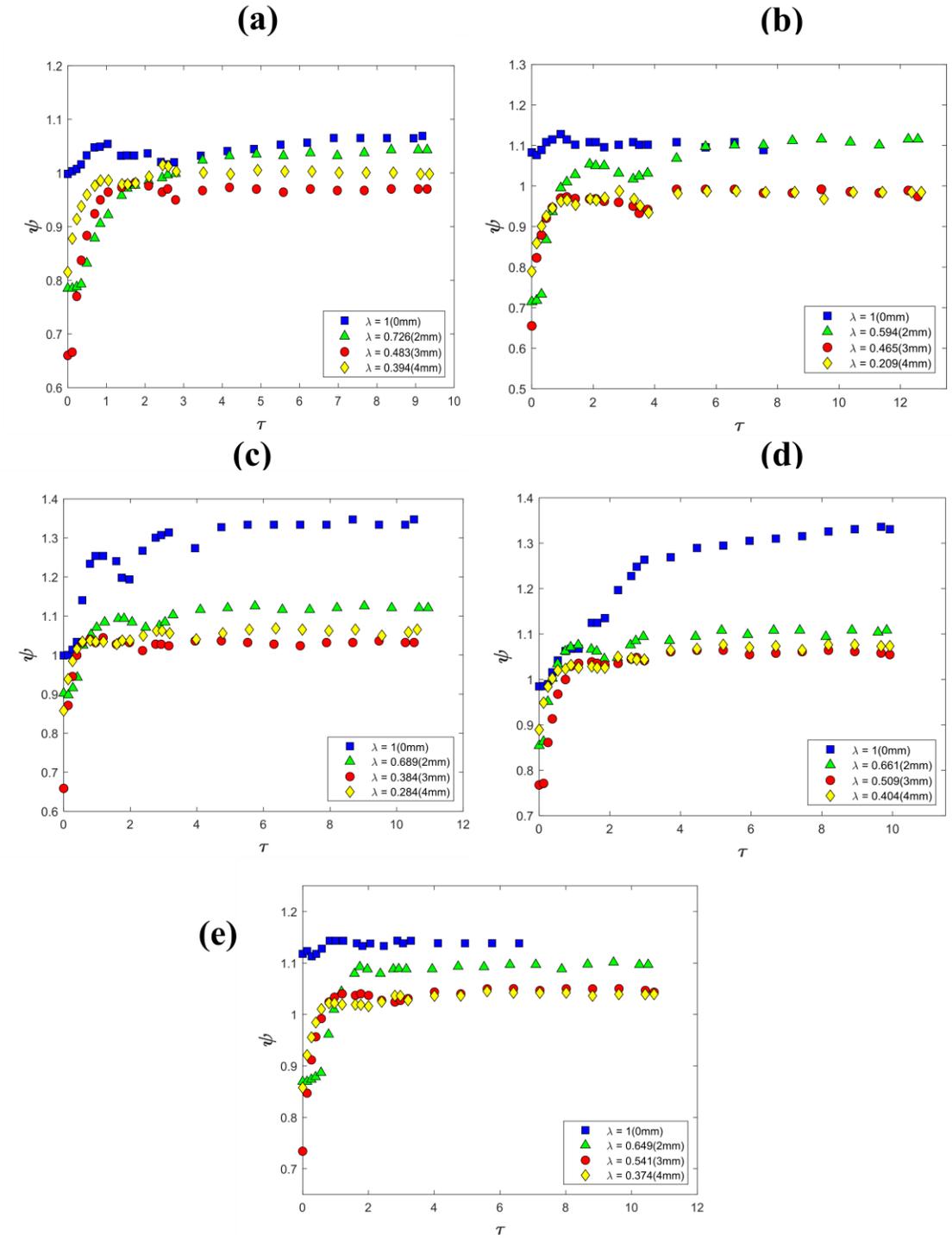

**Figure 7:** Non-dimensional spread length (ψ) vs. non-dimensional time (τ) at various offsets (λ) at 0.4 m/sec velocity for (a) Water drop falling on water drop (b) 0.25 CMC aqueous solution of SDS on water drop (c) 0.5 CMC aqueous solution of SDS on water drop (d) 0.75 CMC aqueous SDS solution on water drop (e) 1 CMC aqueous solution of SDS on water drop

droplet gets squeezed the spreading of the sessile drop progresses till $\tau = 1.033$, the air layer doesn't get drained out and the air pressure causes the falling droplet to lift which causes the sessile droplet to retract also the surface tension force helps in the build-up. The falling droplet gets coalesced with the sessile one as it tries to move upwards wherein the air layer gets drained out completely and falls back as a compound droplet so after $\tau = 2.75$ we see further spreading mostly due to the capillary effects of the solid and the liquid which relaxes to an equilibrium length and the shape remains as a spherical caplet and that happens to be the maximum length of spreading. A similar trend is observed for the smallest offset impact ($\lambda = 0.762$), here the spreading starts as the falling droplet gets squeezed but the air layer gets drained so no lift-off happens at that offset impact the coalesced droplet then spreads gradually and retraction happens at $\tau = 2.48$ after that the wetting nature of the solid substrate causes the compound droplet to spread to a maximum length which also happens to be the equilibrium length.

As the offset gets increased ($\lambda = 0.483$) the coalescence happens as soon as the droplet lands partially on the static droplet and partially on the solid substrate, The impacting edge spreads outwards and there is minimal impact energy so the non-impacting side stays undisturbed for a while, after the surface wave gets towards the non-impacting edge it starts to feel the momentum disturbance and the left edges moves to a certain extent. The surface tension forces make the compound droplet to recoil at $\tau = 2.09$ and the free surface bulges upwards followed by capillary spreading and pinning of the edges which doesn't lead to spreading but the free surface remains oscillating. The largest offset ($\lambda = 0.394$) impact droplet lands completely on the substrate just touching the edge of static drop hence their edges coalesce upon contact and the little inertia it has gained makes the impacting end spread on the solid surface and the non-impacting end hardly moves, there's a retraction phase at $\tau = 1.05$ then followed by re-spreading just like the previous offset impact and the edges get pinned making the compound droplet to come to a halt.

It is also observed that there could arise a situation when the falling droplet hits concentrically on the sessile drop, the air layer shoehorned between the two droplets doesn't get completely drained even when the hitting droplet starts to retract after getting squeezed into the sessile drop. As a result, the falling droplet gets completely rebounded and bounces off the sessile drop and the two drops don't get coalesced upon impact, the head-on impacts of Fig. 7(b) and (e) are indicative of that. Fig 7(b) shows the dimensionless spread length over dimensionless time at various offsets of a 0.25 CMC aqueous solution of SDS falling on a sessile water drop, the head-on impact shows a jump in spreading length at $\tau = 2.15$ as because the falling droplet lands on the substrate after getting bounced off and merges with the sessile drop after spreading individually. The smaller offset ($\lambda = 0.594$) shows similar spreading behavior as the previous impact case there is a $\tau = 2.35$ followed by re-spreading to an equilibrium which is the maximum spreading length. The larger offsets ($\lambda = 0.456, 0.209$) almost show similar spreading lengths as the inertial spreading is less prevalent and the maximum spreading is by the wetting nature of the substrate and the maximum spreading length is almost equal as it reaches for a stable configuration.

Fig. 7(c) shows the spreading nature of 0.5 CMC aqueous SDS solution colliding with a water drop. The axisymmetric impact shows an increase of the spreading length in the inertial regime as the falling droplet pushes the sessile drop to a pancake shape and due to the air layer draining off , the two droplet coalesces as their free surface comes into contact. The increase in spreading is halted by sudden retraction at $\tau = 1.18$ owing to the surface tension of water then the compound drop re-spreads due to the wetting nature of the substrate which overcomes the cohesive pull and reaches a peak where the spreading halts albeit some free surface oscillations. The smaller offset ($\lambda = 0.689$) follows a similar nature as the head-on impact with minor oscillation, these oscillations amplitude gets smaller with an increase of offset distance. The falling drop slides past the sessile droplet and gets coalesced pushing out the right edge at 3mm offset ($\lambda = 0.384$), the non-impacting end remains unmoved which also happens to be the same for the 4 mm offset (0.284). After an initial spreading of the impacting edge over the substrate there are interplay of the adhesive and cohesive forces which leads to minor retraction and re-spreading finally reaching out for an equilibrium position.

Further lowered surface tension fluid which is 0.75 CMC of aqueous solution of SDS is allowed to fall onto the static water drop and the corresponding graphical representation of spreading is shown in Fig. 7(d). It is observed that the concentric impact spreading peaks upwards after an initial spread and a minor plateau. As the impact happens the falling droplet gets squeezed onto the sessile one there arises a necking phenomenon and a capillary column gets formed pinching off a tiny daughter droplet upwards and due to the large surface tension difference, the Marangoni spread[38] and subsequently with the strong capillary force causes the combined drop to spread continuously. The smaller offset shows a similar trend however there is a retracting phase at $\tau = 1.11$ and then re-spreads with minor oscillations. Just like it was observed in the previous case of 0.5 CMC the larger offset impacts ($\lambda = 0.509, 0.404$) in this case show minor wavering could still be observed but of lesser amplitude.

Fig 7(e) depicts the spreading characteristics against time for a 1 CMC aqueous SDS solution falling on a sessile water drop at various offsets. As discussed earlier in this section due to the incomplete draining of the air layer sandwiched between the droplets creates a pressure that propels the falling droplet upwards to bounce it off the sessile drop, a similar incident is depicted in the head-on collision here there is a jump from $\tau = 6.58$ to $7.04$ the additional spreading is due to the reason that the falling droplet after bouncing off gains inertia as it coalesces and spreads along with the sessile drop. The smallest offset impact at 2mm ($\lambda = 0.649$) shows the same squeezing action which makes the initial spread as the falling droplet hits the sessile one and due to the complete draining of the air layer between them, the drops coalesce as the falling droplet pushes the impacting edge downwards. The surface waves travel towards the non-impacting edge and make that edge move which leads to a steep increase in spreading till a point of retraction at $\tau = 1.81$, the surface tension pulls back only to a minor extent due to the lowest surface tension force after that there's re-spreading oscillating between the capillary force due to wettability of the substrate and the weak surface tension inward pull. In $\lambda = 0.541$ offset the spreading curve steeps upwards due to the sliding down of the falling droplet as it hits on the furthest edge of the static drop and the right edge of the merged drop spreads on the substrate due to the lower surface tension and hydrophilic nature of the substrate. The non-impacting edge gets into motion very late due to lower inertial strength and the resisting surface tension of the pure water drop, after a minor retraction at $\tau = 1.87$ the compound drop spreads further with minor oscillation. The furthest offset $\lambda = 0.374$ shows a similar nature of spreading as the former offset there is less difference between the two since the falling droplet falls off the solid substrate completely and spreads like an impact of a single drop on a solid substrate and the former offset impact so happens that the falling drop slides past the impacting edge and lands on the substrate due to the greater surface tension force on the sessile water drop.

*c. Dynamics at impact velocity of 1.6 m/s*

The impact phenomena at a higher velocity of 1.6 m/sec are recorded as the falling drop of varied surface tension hits a static water drop head-on and at increasing offsets. Weber number is of the range 184-615, Capillary number varies from 0.2-0.7 and Reynolds number is from 7300-8528. Figure 8(a-e) captures the non-dimensional spreading length against dimensionless time at this velocity. Due to higher velocity, the inertia effect is dominant and we see that there is an immediate peak in spreading succeeded by a smoother retraction phase and plateau in water on the water impact scene whereas when the falling droplet is of reduced surface tension the retraction is minimal as the surface tension forces are unable to pull back after inertia depletes.

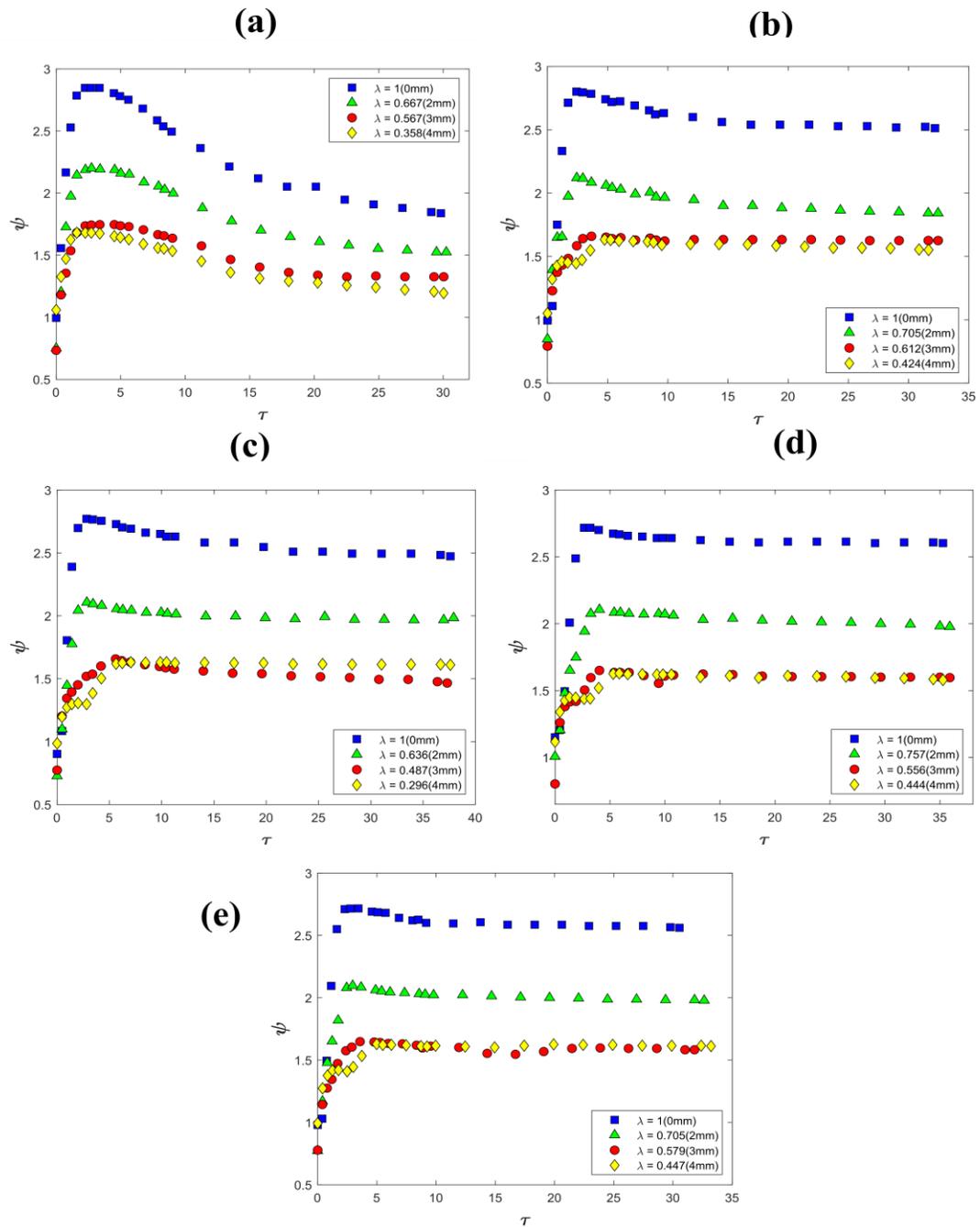

**Figure 8:** Non-dimensional spread length (ψ) vs. non-dimensional time (τ) at various offsets (λ) at 1.65 m/sec velocity for (a) Water drop falling on water drop (b) 0.25 CMC aqueous solution of SDS on water drop (c) 0.5 CMC aqueous solution of SDS on water drop (d) 0.75 CMC aqueous SDS solution on water drop (e) 1 CMC aqueous solution of SDS on water drop

Fig 8(a) shows the spreading characteristics of a water droplet falling on another water drop at different offsets. The head-on impact clearly spreads to a greater extent which is apparent since it involves higher inertial impact transmission to the static droplet causing a dent in the center and spreading of the fluid radially and formation of a rim at the outer edge. The maximum spreading is reached at $\tau = 2.24$ and there is a smooth downward sloping in the graphs which shows the pulling back of the edges due to surface tension and the length curve equilibrates after the balance of forces in the contact line. As the offset is increased, we see that the maximum spreading length and the equilibrium lengths are decreased which is perceptible because at larger offsets the impinging inertia decreases and the static drop resists the momentum more.

Fig. 8(b) depicts the spreading of a 0.25 CMC SDS droplet falling on a static water drop. Compared to the water drop falling on the water drop here after the steep kinematic phase of spreading the plateau phase is less steeper than in the previous case and the equilibrium length of the reduced surface tension droplet impact is greater than the similar surface tension impact case and with decreasing the overlap ratio the maximum and equilibrium lengths are also decreased. The variation of the spreading length over time for a 0.5 CMC aqueous solution of SDS hitting a water droplet is portrayed in Fig. 8(c) and it is observed that there is no specific change in the spreading characteristics for the head-on case the time on maximum spread length is reached is roughly the same as that of 0.25 CMC droplet hitting on water droplet required to reach and as the offset increases the retraction phase is minimal, the lowered surface tension makes the impacting edge recede very less and due to the mixing of the fluid there is also lesser pulling action in the non-impacting edge the largest offset ($\lambda = 0.296$) recedes even less which makes the equilibrium length to me more than that of the previous offset. Fig. 8(d) depicts the spreading length variation of further lowered surface tension of 0.75 CMC SDS solution droplet falling on static water drop concentrically and at other non-concentric variations, here the spreading nature remains the same as the previous case the plateau region is almost vanished and the spreading length varies only to a miniscule amount after initial kinematic regime, the larger offsets ($\lambda = 0.556, 0.444$) doesn't show much difference in their spread length due to the higher impact energy which makes the farthest offset impact to spread more due to the substrate wettability.

Lastly, the lowest surface tension droplet which is 1 CMC of SDS solution hits a sessile water drop at the maximum velocity of 1.64 m/sec, the spreading characteristics are shown in Fig. 8(e). As for the head-on impact, there is a shorter plateau period after the inertial phase and the fluid reaches an equilibrium length that doesn't change further, the smaller offset's dynamics are the same as the previous one which shows no retraction after the initial growth phase and as the offset is increased the nature of the dynamics are unchanged as the previous ones.

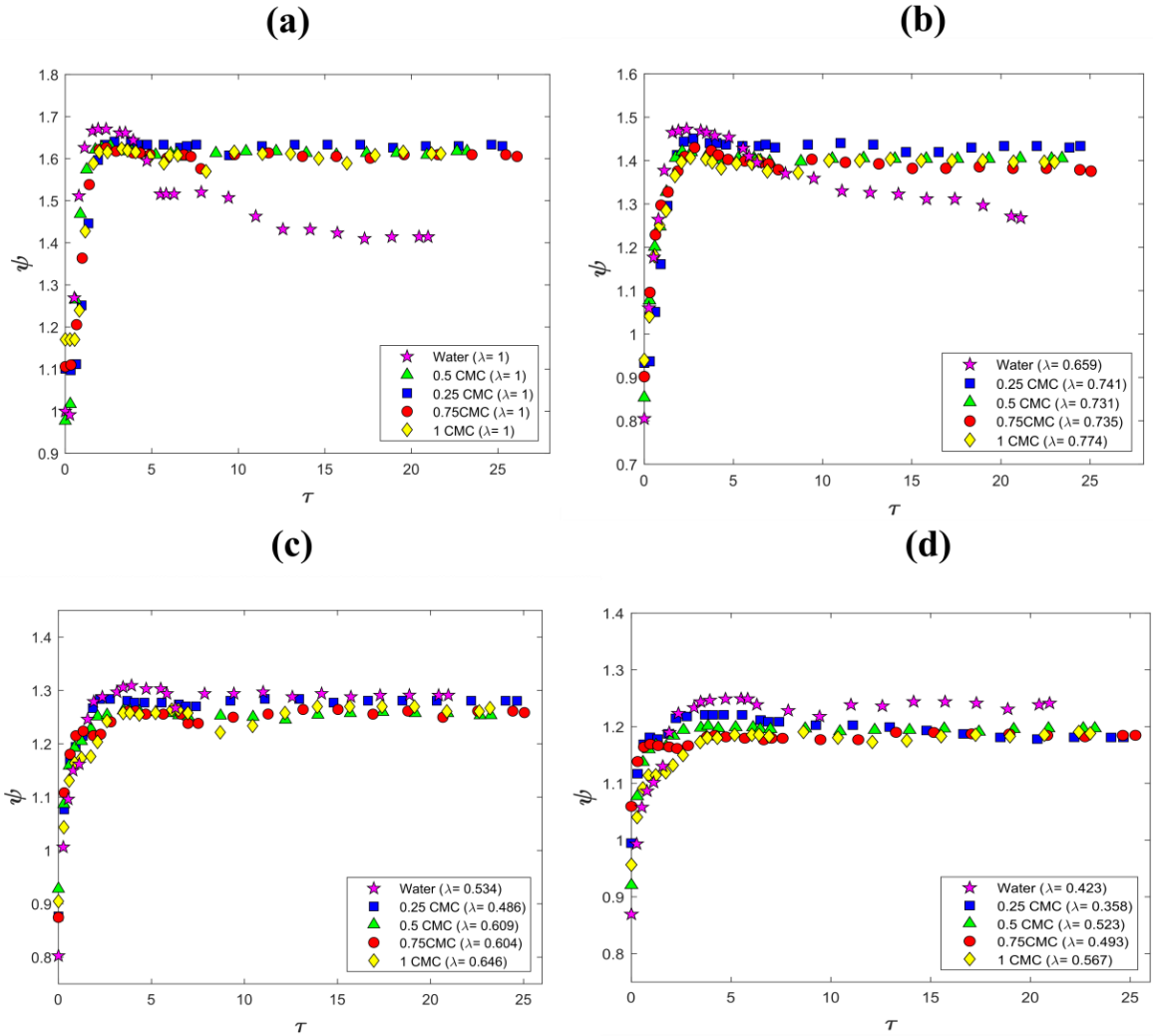

**Figure 9: Comparison of spreading length for impacting drops of different fluid type at a particular offset for falling droplet velocity of 1 m/sec (a) head-on impact (b) impact at 2 mm offset (c) impact at 3 mm offset (d) impact at 4 mm offset**

The comparison of the spreading lengths for all the droplet impacts of the different fluids at the same offset for the impact velocity of 1 m/sec has been depicted graphically in Figure 9 (a-d). The concentric impacts are shown in Fig. 9(a) where the impacting fluid droplet ranges from water (same surface tension as the sessile drop) to 1 CMC aqueous SDS solution (lowest surface tension difference between the falling and the sessile drop). The maximum spreading lengths reached the impact velocity of 1 m/sec are roughly the same for impacting drops of varied surface tension although from the figure it can be inferred that it is maximum for the water droplet, it is because the size of the water droplet is more compared to the other droplets and there is a significant amount of receding that is observed in the case of the water drop hitting on another water drop whereas a small difference in lowering surface tension makes the receding regime diminish. As the impact offset is increased the receding phase starts to taper out for the water-on-water impacts (see Fig. 9(b-d)) because the non-impacting end gets a lesser amount of momentum disturbance so the recoil is less from that end leading to lesser retraction altogether. It is observed that the time required for reaching the maximum spread length increases subtly as the surface tension of the falling droplet is lowered for concentric impacts and smaller impact offsets (2mm) however for the larger offsets (3mm, 4 mm) the spreading time is significantly larger for the lower surface tension drops (0.75, 1 CMC) to reach the maximum spread owing to spreading of the impacting edges (Fig. 9(c-d)) as a result of the hydrophilic nature of the substrate which makes the larger offsets take more time to reach the maximum spread length. With the increase in offsets, the equilibrium lengths of the impact of water on water drop reach to the close vicinity of the equilibrium lengths of the other impacting drops of lower surface tension.

Figure S7 (a-d) (refer Supplementary material) shows the comparison of the spreading lengths at a particular offset for the falling droplet velocity of 0.4 m/sec for falling droplets of various fluids. At such a low velocity the surface tension is dominant and it is observed that after an initial kinetic regime spreading, the rest of the spreading is of capillary spreading regime and hence the lower surface tension droplet impact shows a more capillary-driven spreading than the water drop on the water drop. The maximum spreading length keeps on increasing till the equilibrium is reached so in this case the equilibrium spreading length is considered as the maximum spreading length. Fig. S7 (b) shows the spreading lengths at a smaller (2mm) offset, we see that the receding nature of the water-on-water drop is minimal and it creeps further due to the substrate wettability the other lower surface tension drop impact shows more oscillations and spreads further than the water drop impact. With a further increase in offset distance (3mm) shown in Fig. S7(c) the equal surface tension coalesced drops show some oscillations with decreasing amplitude and the nature of the other lower surface tension impacts also follows the same spreading nature. The largest offset (4mm) impact is depicted in Fig. S7(d) the water drop falling on the water drop shows oscillations with smaller amplitude but the lower surface tension droplet impacts show greater amplitude oscillations than the water drop and the spreading length increases as well. It is a consequence of the higher spreading nature of a low surface tension droplet and the hydrophilic nature of the substrate because at the largest offset the falling drop lands on the solid substrate and then spreads and touches the sessile droplet while the impacting edge continues to spread further like a single drop.

The variation of the spreading lengths against non-dimensional time of all the fluid droplets at a particular offset for the falling velocity of 1.65 m/sec is plotted in Figure S8 (a-d) (Supplementary material). Due to high velocity, there is a steep rise in inertial spread for all the offsets and all the different surface tension liquid drop impacts. The water drop on water impact shows the highest retraction due to the high surface tension of water the pulling action is prominent. In Fig. S8(a) which shows the head-on collision for all the varied surface tension liquid impacts, due to a slightly larger size of the water droplet, the maximum spread length is more in that respect than the lower surface tension droplet the time required for reaching the maximum spread length is roughly the same for all the different surface tension liquid droplets. It is noteworthy that even for surface tension difference of 0.25 CMC between the falling droplet and the static one reduces the retraction regime significantly for all the offset impacts, whereas the retraction regime of water drop impact on water drop impact reduces with increasing offset. Fig. S8(b) shows the spreading dynamics at a 2 mm offset for all the different types of fluids, here the time taken for all the different fluids is also the same and the water-on-water drop impact shows a reduced retraction. As the offset is increased further (3-4 mm, Fig. S8(c-d)) it can be seen that the time at maximum spreading increases for the lower surface tension droplet impact it can be attributed to the fact that on higher offset impacts the disturbance wave reach slower through the static water drop of higher surface tension to displace the non-impacting edge resulting in reaching towards the maximum spreading length a little bit slower, this effect is more pronounced in the 4mm offset impact (Fig. S8 (d)).

The effects of the impact velocities on the spreading length for the different fluid droplet impact at a particular offset have been shown in Figure 10(a-e). For the case of a water drop hitting on a water drop, the spreading length variations at different velocities are presented in Fig. 10(a) and it is evident that the impact velocity is directly proportional to the top kinematic spreading regime also the retraction is prominent in higher energy impacts bringing the equilibrium lengths lesser than the maximum spreading length. In lesser energy impacts the receding is less because of less energy and the equilibrium lengths either reach the level of the maximum spreading lengths or go a little bit beyond that which is due to the capillary spreading. Upon reducing the surface tension of the falling droplet (0.25, 0.5, 0.75 1 CMC) we see a departure from the spreading

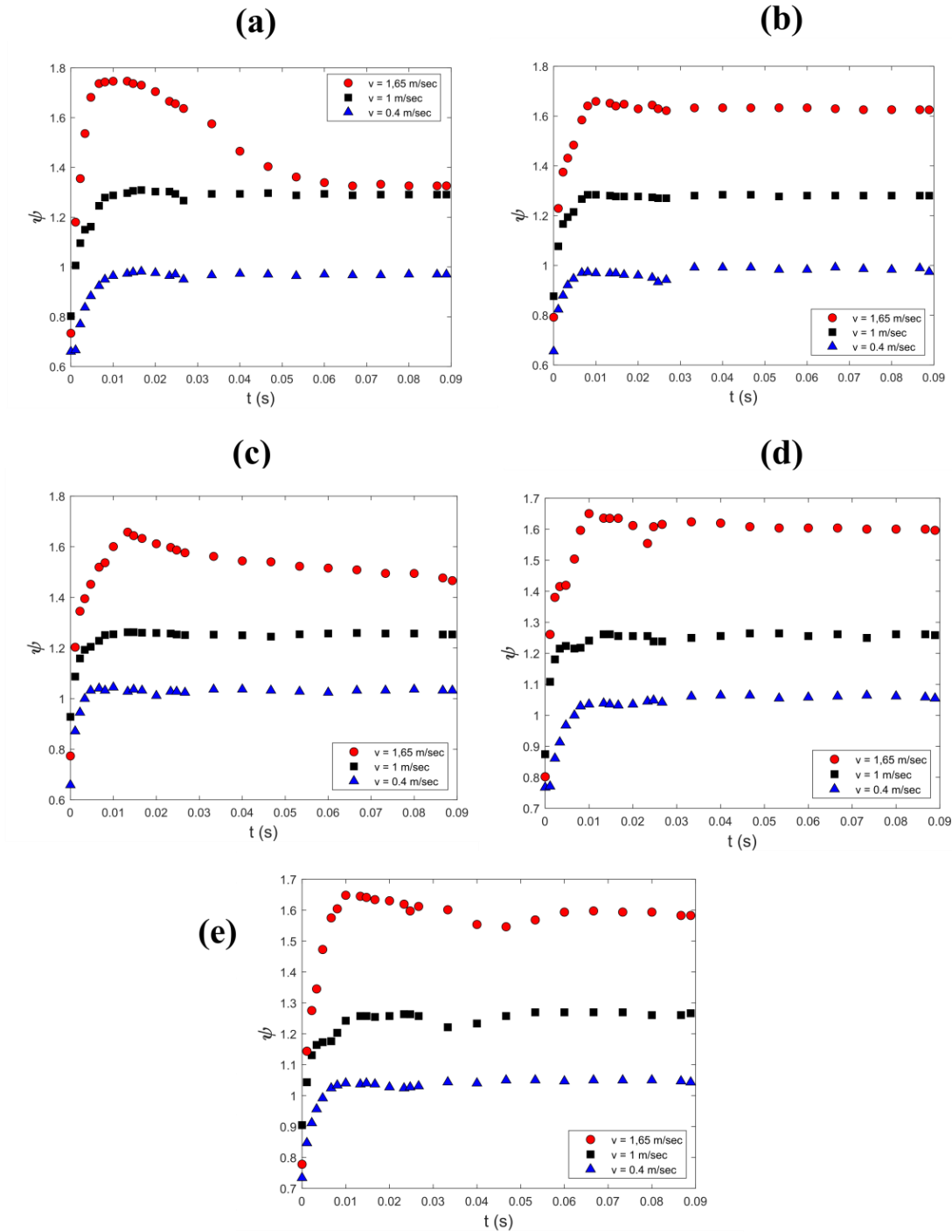

**Figure 10: Comparison of spreading length with fluid type for a particular offset for different velocities (a) water drop on water droplet (b) 0.25 CMC drop on water drop (c) 0.5 CMC drop on water drop (d) 0.75 CMC drop on water drop (e) 1 CMC drop on water drop**

dynamics of a high surface tension falling droplet (water), at the highest velocity a 0.25 CMC droplet falling on a water droplet shows very less receding phase as it is shown in Fig. 10(b) also there is very less spreading after the maximum spreading length is reached.

A similar behavior is shown at a moderate velocity of 1 m/sec and a lower velocity of 0.4 m/sec it is seen that after initial receding the compound droplet creeps upwards due to capillary spreading. Fig 10(c) shows the spreading lengths of a 0.5 CMC aqueous SDS droplet falling on a sessile water drop against time at different velocities, we see a retraction after kinematic spreading regime on the spreading at the highest velocity impact and it is mainly the retraction of the non-impacting edge, at lower velocities (1, 0.4 m/sec) the time required for maximum spreading is less than at the highest velocity which is evident because higher velocity is indicative of higher inertial energy which enables the combined drop to spread more. Viscous dissipation and some oscillations are seen on the highest velocity impact (1.65 m/sec) of a 0.75 CMC droplet with a static water drop as shown in Fig. 10(d) and a similar nature is seen on the 1 m/sec impact spreading at the lowest velocity the capillary spreading is prominent after a minor pullback after the initial kinematic regime is observed. Quite interestingly when the surface tension of the impacting droplet is lowered to 1 CMC, the spreading nature at the highest velocity (1.65 m/sec) shows a retraction and capillary spreading even at that velocity which is due to lower surface tension and hydrophilic nature of the substrate as depicted in Fig. 10(e). At the velocity of 1 m/sec spreading and viscous dissipation are seen after the initial kinematic phase and the rest of the regime is quite similar to the spreading at the highest velocity (1.65 m/sec), the lowest impact velocity (0.4 m/sec) however shows the similar capillary spreading nature after a slight retraction phase.

*d. Measure of drawback and maximum spread*

The contrast between the maximum spreading length ($\psi_{max}$) and minimum spreading length ($\psi_{min}$) is the measure of the extent of pulling away of the impacting droplet towards the first/stationary droplet which is termed as "drawback"[30, 35] and it is undesired and should be kept to a minimum value for the case where a steady line printing is required otherwise the drawback could lead to break-up in the line. Drawback arises mainly due to the domination of the surface tension force which pulls the outward edges inwards after reaching the maximum spread after the inertial regime.

Figure 11(a-e) illustrates the drawback which is $\Delta\psi$ ($\Delta\psi = \psi_{max} - \psi_{min}$) against the offset ($\lambda$) for the various surface tension falling droplets. It is apparent that the highest surface tension of the falling droplet in our case is a water drop falling on another water drop hence we see that from Fig. 11(a) that the drawback for a head–on collision is for the highest velocity (1.4 m/sec) is the maximum that is what the experiments show and decreases as the offset is increased also with a

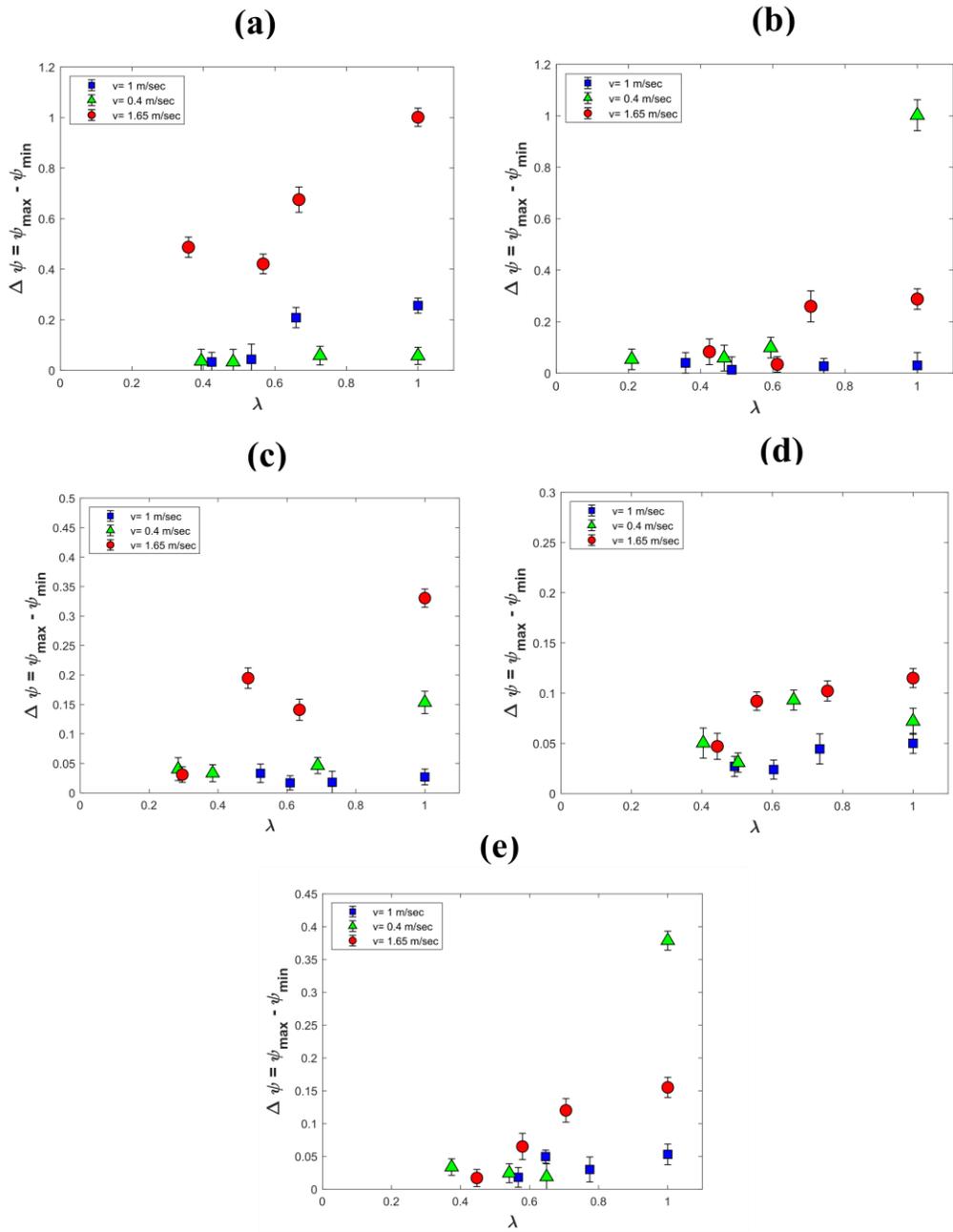

**Figure 11: The change between maximum spreading length ($\psi_{max}$) and minimum spreading length ($\psi_{min}$) which is also known as "drawback" as a function of offset between the impacting drops for different impacting velocities ($\lambda$) (a) water drop on water droplet (b) 0.25 CMC drop on water drop (c) 0.5 CMC drop on water drop (d) 0.75 CMC drop on water drop (e) 1 CMC drop on water drop**

reduction in impact velocity the drawback also decreases as for the water-on-water case the minimum drawback is least for the lowest velocity (0.4 m/sec) impact case. Fig. 11(b) shows the drawback of a 0.25 CMC aqueous SDS droplet falling on a water drop at various velocities and offsets. Here, for the highest velocity case the drawback is reduced significantly than the water-on-water case, it is to be noted that there is a highest drawback shown for the lowest velocity (0.4 m/sec) at head-on impact which is because of the droplet bouncing since the air layer is not drained as the falling droplet makes contact with the sessile water drop barring that circumstance the rest of the drawback shows a similar trend as the former case. When the surface tension of the falling droplet is reduced to 0.5 CMC the drawback is quite similar to the 0.25 CMC droplet case for the highest velocity (1.65 m/sec) which is depicted in Fig. 11(c), however for the instance of the lowest velocity (0.4 m/sec) there is additional spread because of the capillary spreading so we see an increment of drawback however for the instance of 1 m/sec there is a very little drawback.

Fig. 11(d) shows the drawback for the case of a 0.75 CMC aqueous SDS droplet falling on a sessile water drop and it shows a further reduction in drawback for all the velocity ranges, the reduction of surface tension to that extent plays a part in minimizing drawback since the key competition between inertia and surface tension for the drawback to happen the former component is dominant in the play. The lowest surface tension droplet hitting on the sessile water droplet is a 1 CMC aqueous SDS droplet and the drawback for that case is plotted in Fig 11(e) and it shows a similar level of drawback as the former one which is evident as the surface tension is the least. For the head-on collision at the lowest velocity (0.4 m/sec), there is a bouncing-off occurrence and that is why there is a highest drawback shown in the head-on case similar to the one that happened with the 0.25 CMC droplet hitting the sessile water drop head-on. It can be said that by lowering the surface tension of the falling droplet the drawback can be minimized substantially.

In this study, a regression analysis is done to combine the impact of various factors, such as droplet inertia, liquid surface tension, and droplet offset, on the maximum spreading length of droplets. A single empirical correlation is developed to summarize the maximum spreading length for each droplet, impinging velocity and spacing between the droplets and the result is compared with the previous works of Graham et al. and Li et al. and the results are within the closed range of each of the correlations given in the works of the above-mentioned researchers. In their work, Graham et al. used a special Weber number ($We_{sl}$) however Li et al. used the standard Weber number in their regressed equation, following the latter's work we used the standard Weber number. So, carrying out a least squared fitting of the natural logarithms of $We$, $\lambda$ and $\psi_{max}$ the regression equation comes as:

$$\psi_{max} = 0.65(\lambda)^{-0.33}(We)^{0.2} \tag{4}$$

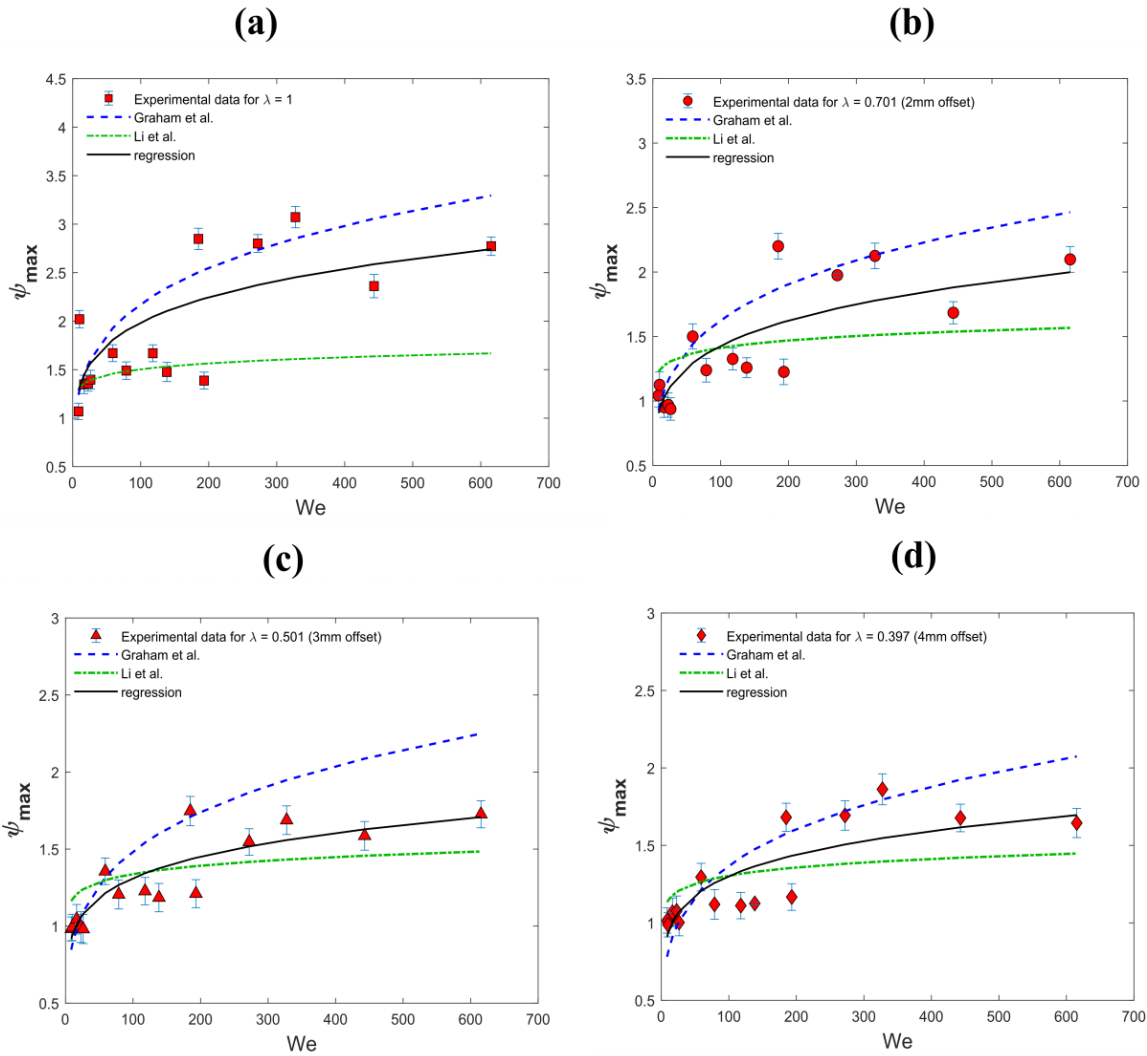

**Figure 12: Comparison between regression curve obtained from equation (4), experimental values of maximum spread length and correlation of previously published results for Maximum spread length ($\psi_{max}$) as a function of Weber number (We) for (a) head-on impact (b) impact at 2 mm offset (c) impact at 3 mm offset (d) impact at 4 mm offset**

The correlation coefficient obtained is $R^2 = 0.81$ which indicates the trend obtained is a consistent one. Figure 12 (a-d) shows the regression model curve along with the experimentally obtained data and shows the comparison between the regression curves obtained by Graham et al. and Li et al. Fig 12 (a) gives the maximum spreading length against Weber numbers for head-on collision. For the rest of the sub-plots, Fig 12 (b-d) shows the trend for the maximum spreading length at 2mm offset, 4 mm offset and 4 mm offset respectively. From the images it can be ascertained that head-on collision has the maximum spreading length and the regression model predicts the trend correctly.

## IV. Conclusions

The post-impact dynamics and analysis of the spreading length of two drops of varied surface tension without changing the viscosities over a hydrophilic surface has been experimentally carried out with the motive of gaining insight into the various parameters viz. droplet surface tension, offset between the falling and sessile drop, the velocity of the falling drop affecting the collision and spreading mechanism. A sessile water drop on a glass surface was allowed to be hit by various drops ranging from water (equal surface tension to that of the sessile drop) to 1 CMC of aqueous solution of Sodium Dodecyl Sulfate (SDS) at various spacing between the droplets and falling at different velocities. At a moderate velocity of 1 m/sec and at the highest velocity of 1.65 m/sec the water drop on water showed the maximum spreading length but there is a significant retraction at head-on and at less offset (2mm) hence the equilibrium length is much lesser than the maximum spreading lengths. However, at higher offsets (3mm, 4 mm) the retraction is reduced to some extent. It is found that reducing the surface tension of the falling water drop by adding some surfactant (Sodium Dodecyl Sulfate) to a minor extent which in our case is 0.25 CMC affected abating the retraction of the coalesced droplet significantly further lowering the surface tension the level of retraction remains the same in some measure. This finding is important for line printing where the "drawback" is not welcomed. When the droplet velocity is the lowest which in our case is 0.4 m/sec, it is found that the maximum spreading length becomes greater as the surface tension of the falling droplet is lowered. The equilibrium spreading length goes on increasing due to the capillary spreading as a result of the wetting nature of the solid substrate and becomes the maximum spreading length which is lowered as the offset between the two droplets increases ($\lambda$ decreases). On head-on collision, there is a chance that the air layer between the two droplets doesn't get drained out as the falling drop makes contact with the sessile one resulting in a complete rebound of the colliding drop. A drawback study is also carried out which shows that the drawback is maximum for the highest surface tension droplet and for the highest velocity and lowest offset. Finally, a regression analysis is done to predict the maximum spreading length for different impact conditions and spacing between the drops and it is compared with existing correlations and the results agree with it.

**Competing interest statement:** The authors declare no conflicts of interest concerning this research.

**Acknowledgements:** PKS thanks NIT Arunachal Pradesh for the doctoral scholarship, and the Multiphase Flow Laboratory, NIT Arunachal Pradesh for access to high-speed camera and other equipment.## References

[1]A.M. Worthington, "On the form assumed by drops of liquids falling vertically on a horizontal plate", Proc. R. Soc. Lond. 25, 261 (1876).

[2]A.L. Yarin, "Drop impact dynamics: Splashing, spreading, receding, bouncing", Annu. Rev. Fluid Mech. 38, 159 (2006).

[3]Y. M. Zhang, Y. Chen, P. Li, and A. T. Male, "Weld deposition-based rapid prototyping: a preliminary study," J. Mater. Process. Technol. 135, 347 (2003)

[4]M. Fang, S. Chandra, and C. B. Park, "Experiments on remelting and solidification of molten metal droplets deposited in vertical columns," J. Manuf. Sci. Eng. 129, 311 (2007)

[5]H. Sirringhaus, T. Kawase, R. H. Friend *et al.*, "High-resolution inkjet printing of all-polymer transistor circuits," Science 290, 2123 (2000).

[6]M. Rein, Phenomena of liquid drop impact on solid and liquid surfaces, Fluid Dyn. Res. 12, 61 (1993).

[7]C. Josserand, S.T. Thoroddsen, "Drop Impact on a Solid Surface". Annu. Rev. Fluid Mech. 48, 365 (2016).

[8]M. Marengo, C. Antonini, I.V. Roisman, C.Tropea, "Drop collisions with simple and complex surfaces", Curr. Opin. Colloid Interface Sci. 16, 292 (2011).

[9]H. Wang, X. Zhu, R. Chen, Q. Liao, B. Ding, "How supercooled superhydrophobic surfaces affect dynamic behaviors of impacting water droplets" Int. J. Heat Mass Transf. 124, 1025 (2018).

[10]Y. Guo, S. Shen, Y. Yang, G. Liang, N. Zhen, "Rebound and spreading during a drop impact on wetted cylinders", Exp. Therm. Fluid Sci. 52, 97 (2013).